\documentclass[12pt]{article}
\usepackage{amsmath,amssymb}
\usepackage{graphicx}
\usepackage{subfigure}
\usepackage[pdftex]{hyperref}
\DeclareGraphicsExtensions{.eps,.bmp,.wmf,.jpg,.pdf}
\numberwithin{equation}{section}
\def\be{\begin{equation}}
\def\ee{\end{equation}}

\def\bea{\begin{eqnarray}}
\def\eea{\end{eqnarray}}

\title{Reconstructing the potentials for the quintessence and tachyon dark energy, from the holographic principle}
\author{L.N. Granda\thanks{ngranda@univalle.edu.co} \\
Department of Physics, Universidad del Valle\\ A.A. 25360, Cali,
Colombia} 
\date{}
\begin{document}
\maketitle

\begin{abstract}
\noindent We propose an holographic quintessence and tachyon models of dark energy. The correspondence between the quintessence and tachyon energy densities with the holographic density, allows the reconstruction of the potentials and the dynamics for the quintessence and tachyon fields, in flat FRW background. The proposed infrared cut-off for the holographic energy density works for two cases of the constant $\alpha$: for $\alpha<1$ we reconstructed the holographic quintessence model in the region before the $\omega=-1$ crossing for the EoS parameter. The cosmological dynamics for $\alpha>1$ was also reconstructed for the holographic quintessence and tachyon models.
\end{abstract}

\section{Introduction}
\noindent Many astrophysical data, such as observations of large scale structure \cite{tegmark}, searches for type Ia supernovae \cite{riess}, and measurements of the cosmic microwave background anisotropy \cite{spergel}, all indicate that the expansion of the universe is undergoing cosmic acceleration at the present time, due to some
kind of negative-pressure form of matter known as dark energy (\cite{copeland},\cite{turner}). Although the cosmological observations suggest that dark energy component is about $2/3$ of the total content of the universe, the nature of the dark energy as well as its cosmological origin remain unknown at present. The simplest candidate for dark energy is the cosmological constant \cite{weinberg}, \cite{padmana}, \cite{sahni} conventionally associated
with the energy of the vacuum with constant energy density and pressure, and an equation of state $w = -1$. The present observational data favor an equation of state for the dark energy with parameter very close to that of the cosmological constant. The next simple model proposed for dark energy is the quintessence ((see \cite{copeland1}, \cite{caldwell}, \cite{zlatev})), an ordinary scalar field
minimally coupled to gravity, with particular potentials that lead to late time inflation. The equation of state for a spatially homogeneous quintessence scalar field satisfies $w > -1$ and therefore can produce accelerated expansion. This field is taken to be extremely light
which is compatible with its homogeneity and avoids the problem with the initial conditions \cite{copeland}. Besides quintessence, a wide variety of scalar-field models have been proposed to explain the nature of the dark energy. These include K-essence models based on scalar field with non-standard kinetic term \cite{armendariz},\cite{chiba}; string theory fundamental scalars known as tachyon \cite{padmana1} and dilaton \cite{gasperini}; scalar field with negative kinetic energy, which provides a solution known as phantom dark energy \cite{caldwell1}. Other proposals
on dark energy include interacting dark energy models \cite{amendola} \cite{yin}, brane-world models \cite{dvali}, \cite{shtanov}, modified theories of gravity known as f(R) gravity, in which dark energy emerges
from the modification of geometry \cite{carroll},\cite{capozziello}, \cite{odintsov}, \cite{sergei1}, and dark energy models involving non-standard equations of state \cite{kamen},\cite{odin} (for a review on above mentioned and other approaches to dark energy, see \cite{copeland}). In all these models of scalar fields, nevertheless, the potential is chosen by hand guided by phenomenological considerations, lacking the theoretical origin.
Another way to the solution of the dark energy problem within the fundamental theory framework, is related with some facts of the quantum gravity theory, known as the holographic principle (\cite{beckenstein, thooft, bousso, cohen, susskind}). This principle emerges as a new paradigm in quantum gravity and was first put forward by t' Hooft \cite{thooft} in the context of black hole physics and later extended by Susskind \cite{susskind} to string theory. According to the holographic principle, the entropy of a system scales not with it's volume, but with it's surface area. In the cosmological context, the holographic principle will set an upper bound on the entropy of the universe \cite{fischler}. In the work \cite{cohen}, it was suggested that in quantum field theory a short distance cut-off is related to a long distance cut-off (infra-red cut-off $L$) due to the limit set by formation of a black hole, namely, if is the quantum zero-point energy density caused by a short distance cut-off, the total energy in a region of size $L$ should not exceed the mass of a black hole of the same size, thus $L^3\rho_\Lambda\leq LM_p^2$. Applied to the dark energy issue, if we take the whole universe into account, then the vacuum energy related to this holographic principle is viewed as dark energy, usually called holographic dark energy \cite{cohen} \cite{hsu}, \cite{li}. The largest $L$ allowed is the one saturating this inequality so that we get the holographic dark energy density.
\begin{equation}\label{eq1}
\rho_\Lambda=3c^2M_p^2L^{-2}
\end{equation}
where $c^2$ is a numerical constant and $M_p^{-2}=8\pi G$.\\
In the work \cite{li} it was pointed out that the infra-red cutoff $L$ should be given by the future event horizon of the universe, in order to provide the EoS parameter necessarily for the accelerated expansion. 
Viewing the scalar field dark energy models as an effective description of the
underlying theory of dark energy, and considering the holographic vacuum energy scenario as pointing in the direction of the underlying theory of dark energy, it is interesting to study how the scalar field models can be used to describe the holographic energy density as effective theories. The holographic tachyon  have been discussed in \cite{setare1}, \cite{jingfei}, the holographic phantom quintessence and Chaplygin gas models have been discussed in \cite{setare2,setare3} respectively. In all this models, the infra-red cut-off given by the future event horizon has been used for the reconstruction of the corresponding potentials. However this cut-off enters in conflict with the causality \cite{li}. Other reconstructing techniques in theories with a single or multiple scalar fields has been worked in \cite{sodintsov}.

In this paper we are interested in how the scalar field models of quintessence and tachyon can be used to describe the holographic scenario proposed in \cite{granda} with an infrared cut-off given by Eq. \ref{eq2} below, which avoids conflict with the causality. Once we fix the constants $\alpha$ and $\beta$ as indicated in \cite{granda}, according to the observational data, we proceed to the reconstruction of the potentials for the quintessence and tachyon fields. To this end we will use the IR cut-off for the holographic dark energy density
\begin{equation}\label{eq2}
\rho_{\Lambda}=3M_p^2\left(\alpha H^2+\beta \dot{H}\right)
\end{equation}
where $H=\dot{a}/a$ is the Hubble parameter and $\alpha$ and $\beta$ are constants which must satisfy the restrictions imposed by the current observational data. Besides the fact that the underlying origin of the holographic dark energy is still unknown, the inclusion of the time derivative of the Hubble parameter may be expected as this term appears in the curvature scalar (see \cite{gao}), and has the correct dimension. This kind of density may appear as the simplest case of more general $f(H,\dot{H})$ holographic density in the FRW background. This proposal also avoids the coincidence problem as the expression for the holographic density contains two terms which track dark matter and radiation as will be clarified.

\section{The model}

\noindent starting from the Eq. (\ref{eq2}), we write the Friedman equation 
\begin{equation}\label{eq3}
H^2=\frac{1}{3M_p^2}(\rho_m+\rho_r)+\alpha H^2+\beta \dot{H}
\end{equation}
where $M_p=(8\pi G)^{-1/2}$ is the Planck mass and $\rho_m$, $\rho_r$ terms are the contributions of
non-relativistic matter and radiation, respectively. This equation can be rewritten in the form (for details see \cite{granda})
\begin{equation}\label{eq4}
\tilde{H}^2=\Omega_{m0}e^{-3x}+\Omega_{r0}e^{-4x}+\alpha \tilde{H}^2+\frac{\beta}{2}\frac{d\tilde{H}^2}{dx}
\end{equation}
in this equation $x=\ln{a}$ and $\tilde{H}=H/H_0$, and the subscript $0$ represents the value of a quantity at present ($z=0$). Solving Eq. (\ref{eq4}), we obtain
\begin{equation}\label{eq5}
\begin{aligned}
\tilde{H}^2=&\Omega_{m0}e^{-3x}+\Omega_{r0}e^{-4x}+\frac{3\beta-2\alpha}{2\alpha-3\beta-2}\Omega_{m0}e^{-3x}\\
&+\frac{2\beta-\alpha}{\alpha-2\beta-1}\Omega_{r0}e^{-4x}+Ce^{-2x(\alpha-1)/\beta}
\end{aligned}
\end{equation}
where $C$ is an integration constant. Using the redshift relation $1+z=a_0/a$ with $a_0=1$, the equation (\ref{eq5}) takes the form
\begin{equation}\label{eq6}
\begin{aligned}
\tilde{H}(z)^2=&\Omega_{m0}(1+z)^3+\Omega_{r0}(1+z)^4+\frac{3\beta-2\alpha}{2\alpha-3\beta-2}\Omega_{m0}(1+z)^3\\
&+\frac{2\beta-\alpha}{\alpha-2\beta-1}\Omega_{r0}(1+z)^4+C(1+z)^{2(\alpha-1)/\beta}
\end{aligned}
\end{equation}
the last three terms in \ref{eq6} give the scaled dark
energy density $\tilde{\rho}_{\Lambda}=\frac{\rho_{\Lambda}}{3M_p^2H_0^2}$
\begin{equation}\label{eq7}
\tilde{\rho}_{\Lambda}=\frac{3\beta-2\alpha}{2\alpha-3\beta-2}\Omega_{m0}(1+z)^3\\
+\frac{2\beta-\alpha}{\alpha-2\beta-1}\Omega_{r0}(1+z)^4+C(1+z)^{2(\alpha-1)/\beta}
\end{equation}
and the corresponding pressure density $\tilde{p}_{\Lambda}$ is obtained from the conservation equation $\tilde{p}_{\Lambda}=-\tilde{\rho}_{\Lambda}-1/3 d\tilde{\rho}_{\Lambda}/dx$ and is given by
\begin{equation}\label{eq8}
\tilde{p}_{\Lambda}=\frac{2\alpha-3\beta-2}{3\beta}\,C(1+z)^{2(\alpha-1)/\beta}+\frac{2\beta-\alpha}{3(\alpha-2\beta-1)}\,
\Omega_{r0}(1+z)^4
\end{equation}
Considering the equation of state for the present epoch (i.e. at z=0) values of the density
and pressure of the dark energy $\tilde{p}_{\Lambda0}=\omega_0\Omega_{\Lambda0}$ and the Eq. (\ref{eq6}) at the present epoch, we obtain the two equations 
\begin{equation}\label{eq9}
\Omega_{\Lambda}=\frac{3\beta-2\alpha}{2\alpha-3\beta-2}\Omega_{m0}
+\frac{2\beta-\alpha}{\alpha-2\beta-1}\Omega_{r0}+C
\end{equation}
and
\begin{equation}\label{eq10}
\omega_0\Omega_{\Lambda0}=\frac{2\alpha-3\beta-2}{3\beta}\,C+\frac{2\beta-\alpha}{3(\alpha-2\beta-1)}\,
\Omega_{r0}
\end{equation}
from Eqs. \ref{eq9}, \ref{eq10} we can write the constants $\alpha$ and $C$ in terms of  $\beta$, with appropriate values for the parameters $\Omega_{m0}$, $\Omega_{r0}$, $\Omega_{\Lambda 0}$ and $\omega_0$. Once $\tilde{\rho}_{\Lambda}$ and $\tilde{p}_{\Lambda}$ are defined, we can write the expression for the deceleration parameter in terms of the constant $\beta$ (see \cite{granda}). Then we select those values of $\beta$ that give the desired redshift transition according to the astrophysical data . In table I we resume the different values found for this constants as used in \cite{granda}. Nevertheless in all this cases the constant $\alpha<1$, which gives a negative power-law in the last term in the expression for the holographic density (\ref{eq7}), allowing values of the EoS parameter $w_{\Lambda}$ crossing the phantom barrier and giving rise to a future "Big Rip" singularity, as is seen in Fig.1 (left) for the dark energy density. Although this is not the typical behavior of the quintessence or tachyon models, we will reconstruct the potentials in the region before the crossing $\omega_{\Lambda}=-1$, i.e. for $z>0$. 
\begin{center}\renewcommand{\tabcolsep}{1cm}
\begin{tabular}{|c|c|c|c|}\hline
\multicolumn{4}{|c|}{$\Omega_{m0}=0.27$\ \ \ $\Omega_{\Lambda0}=0.73$\ \ \ $\Omega_{r0}=0$\ \ \ $\omega_0=-1$}\\ \hline\hline
$\beta$ & $z_T$ & $\alpha$ & $C$\\ \hline
0.3 & 0.38& 0.85& 0.55\\ \hline
0.5 & 0.59& 0.93& 0.67\\ \hline
0.6 & 0.69& 0.97& 0.7\\ \hline
\end{tabular}
\end{center}
\begin{center}
\it{Table I} 
\end{center}
By other hand, we can also consider another set of values for $\alpha$, $C$ and $\beta$ as giving in table II. With this data we can reconstruct the potentials for the quintessence and tachyon holographic correspondence, in the allowed region of the EoS parameter for this models (i.e. $\omega>-1$). Fig.1 (right) shows the dark energy density in this case. To justify this choice of constants, note that the important observational fact considered here is the correct red-shift transition and the adequate total EoS parameter giving rise to accelerated expansion. If we redefine the present value of the holographic EoS parameter $w_0$, we obtain for instance for $w_0=-0.9$, consistent behavior for the deceleration and the EoS parameters as shown in Fig2 (right) and Fig.3 (right) respectively. The value $w_0\approx-0.9$ is within the limits set by the different sources of astrophysical data on the dark energy EoS parameter \cite{steen,carolina,tegmark1,turner}.
\begin{center}\renewcommand{\tabcolsep}{1cm}
\begin{tabular}{|c|c|c|c|}\hline
\multicolumn{4}{|c|}{$\Omega_{m0}=0.27$\ \ \ $\Omega_{\Lambda0}=0.73$\ \ \ $\Omega_{r0}=0$\ \ \ $\omega_0=-0.9$}\\ \hline\hline
$\beta$ & $z_T$ & $\alpha$ & $C$\\ \hline
0.55 & 0.59& 1.01& 0.67\\ \hline
0.65 & 0.68& 1.06& 0.7\\ \hline
0.7 & 0.72& 1.09& 0.72\\ \hline
\end{tabular}
\end{center}
\begin{center}
\it{Table II} 
\end{center}
Note that $\beta$ is the only parameter in this model which needs to be fitted by observational data.

Figures 1 to 3 were constructed with the data given in table I (left figures) and table II (right figures). We see that the red-shift transition also occurs for values of $\alpha$ greater than $1$, avoiding in this case future singularities. 
\begin{figure}[hbtp]
\includegraphics [scale=0.6]{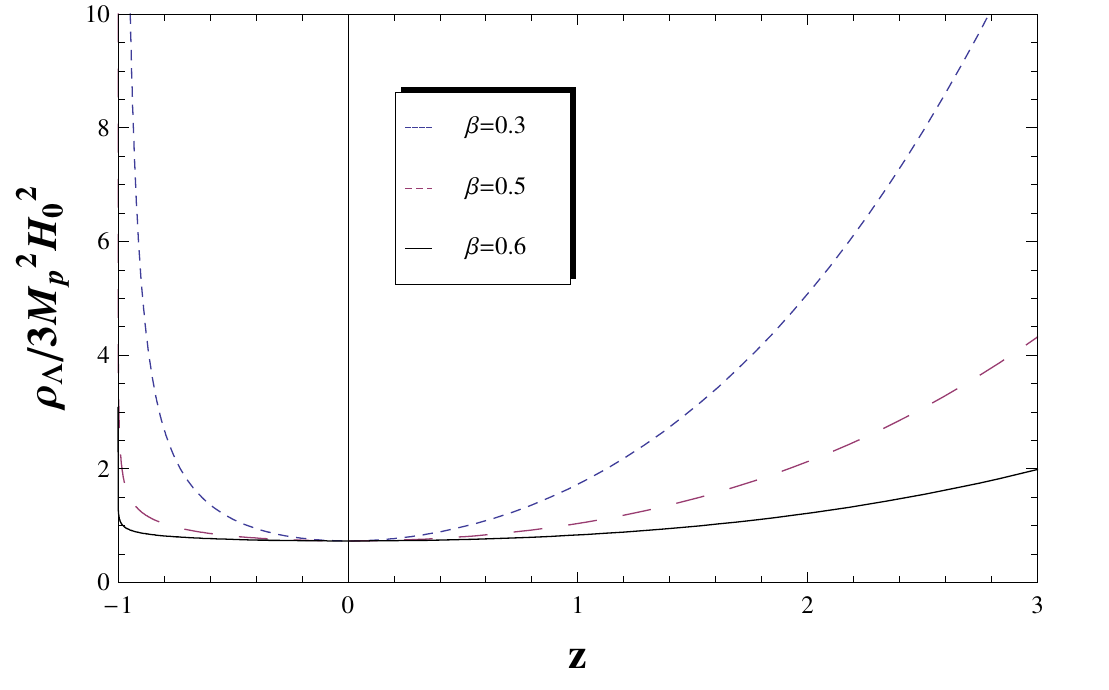}
\includegraphics [scale=0.6]{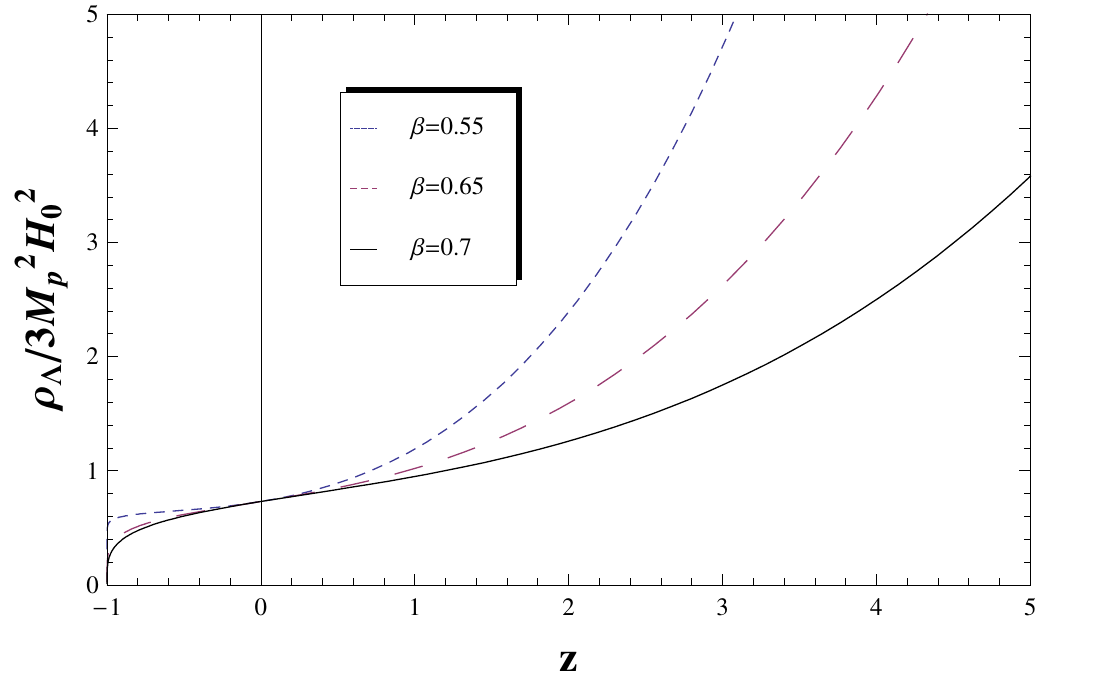}
\caption{The dark energy densities $\tilde{\rho}$ versus redshift, according to table I (left graphic) and table II (right graphic).}
\end{figure}

\begin{figure}[hbtp]
\includegraphics [scale=0.6]{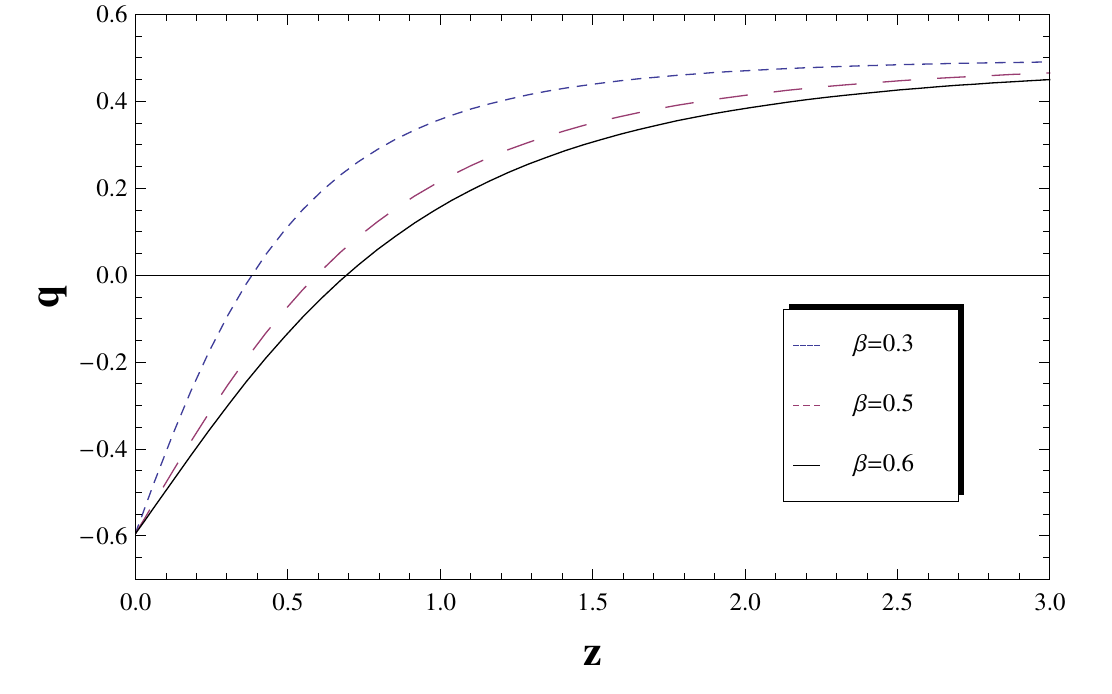}
\includegraphics [scale=0.6]{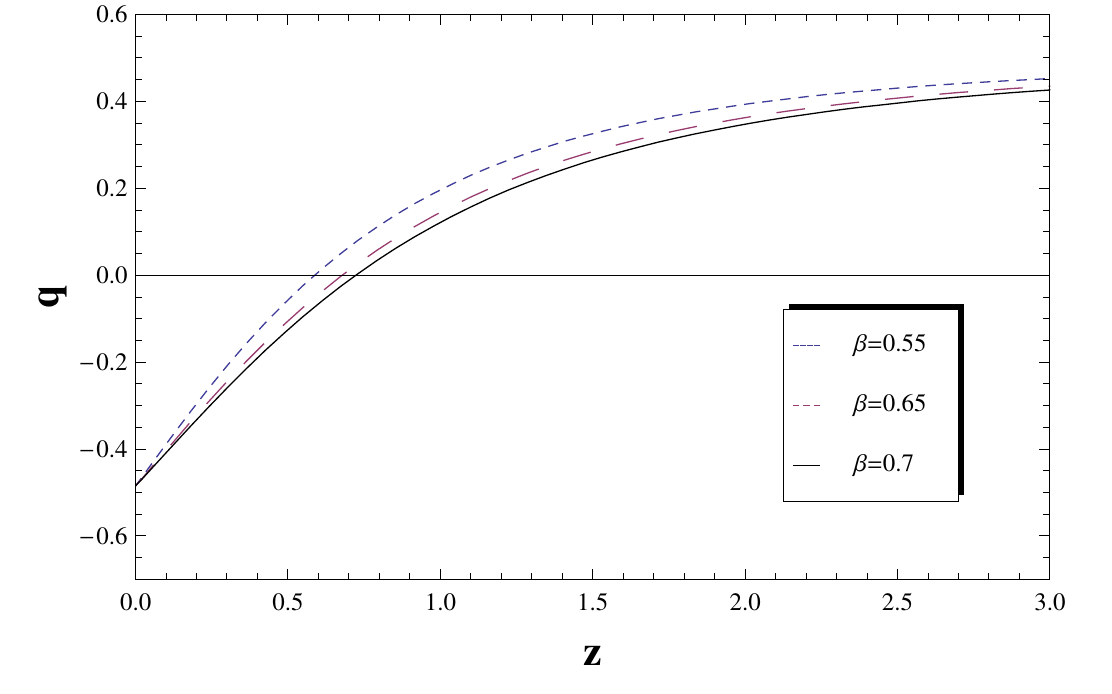}
\caption{The deceleration parameter $q$ versus redshift, according to table I (left), and table II (right).} 
\end{figure}

\begin{figure}[hbtp]
\includegraphics [scale=0.6]{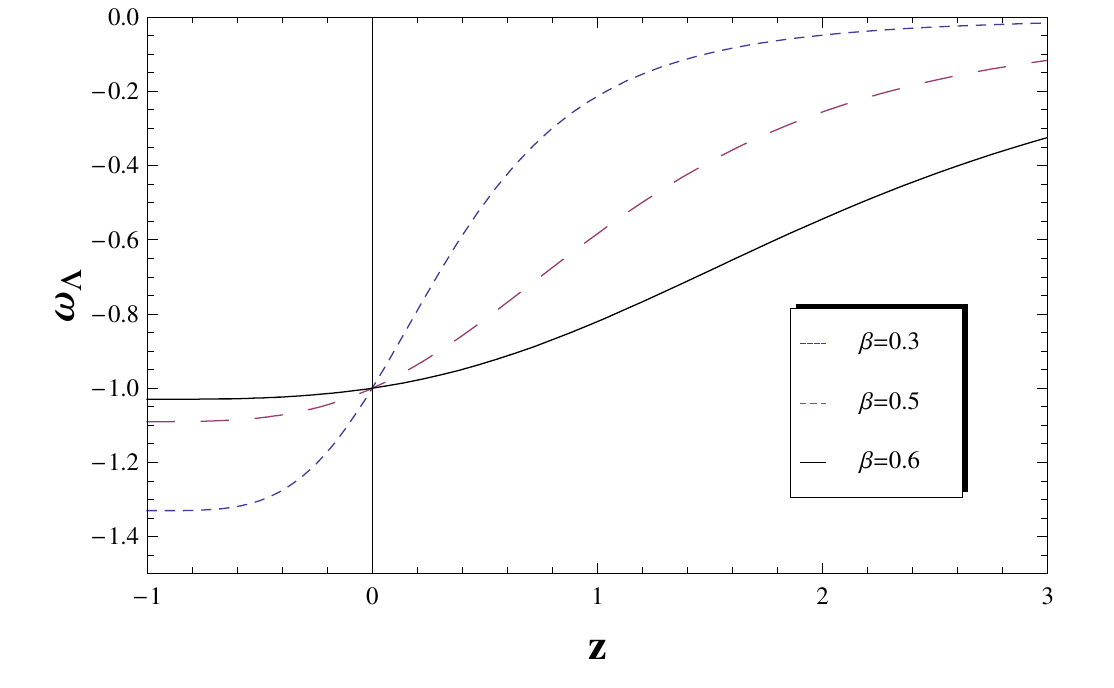}
\includegraphics [scale=0.6]{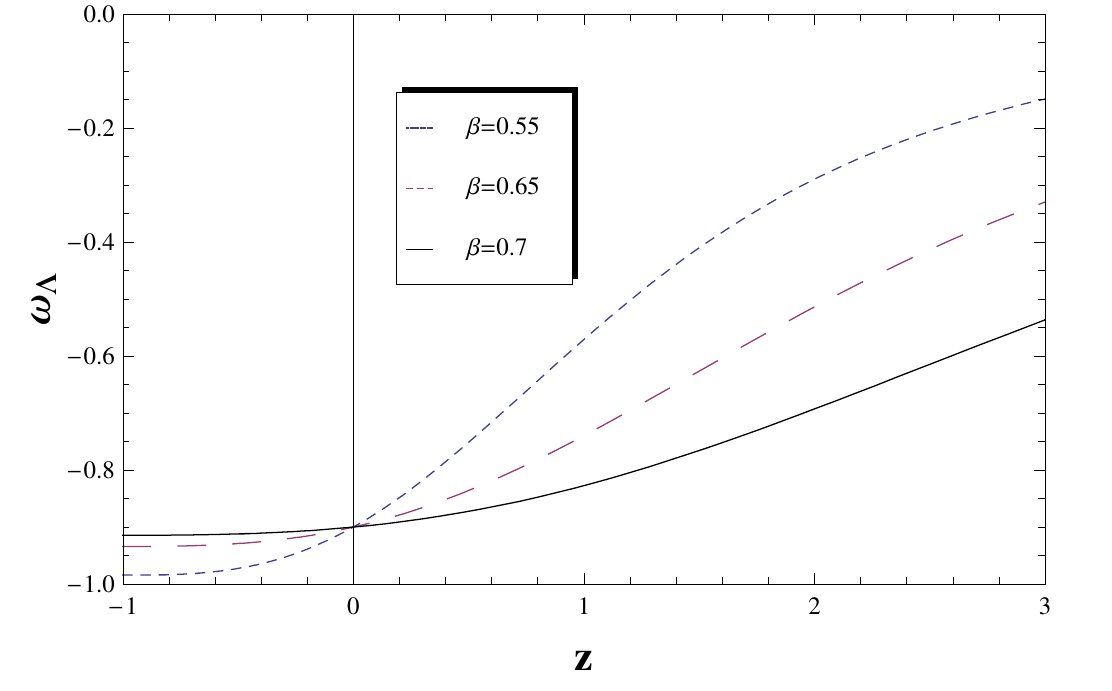}
\caption{The equation of state parameter $\omega_{\Lambda}$ versus redshift, according to table I (left) and table II (right).} 
\end{figure}
Extending the range of $z$ it can be verified that for the values of the constants given in tables I and II, $\omega_{\Lambda}\rightarrow 0$ at high redshift, and the holographic equation of state approaches that of a pressure-less fluid. Therefore, the holographic component becomes sub dominant at earlier times as expected.

\section{Reconstruction of the quintessence model}
\subsection*{$\alpha<1$ case}
In this section, we will discuss the scalar field and potential associated with the quintessence model, and will reconstruct them using the correspondence with the holographic principle.
For the case $\alpha<1$ our holographic model crosses the $\omega=-1$ limit at $z=0$ (see Fig. 3-left). So, there is a region ($w<-1$) forbidden for the quintessence in this case . In order to be consistent with the quintessence model, the holographic quintessence reconstruction will be made for $z>0$.
In the flat FRW background, the energy density and pressure of the quintessence field are given by \cite{copeland}
\begin{equation}\label{eq15}
\rho_{\phi}=\frac{1}{2}\dot{\phi}^2+V(\phi)    \ \ \,\ \ \  p_{\phi}=\frac{1}{2}\dot{\phi}^2-V(\phi)
\end{equation}
which give the equation of state
\begin{equation}\label{eq16}
\omega_{\phi}=\frac{\dot{\phi}^2-2V(\phi)}{\dot{\phi}^2+2V(\phi)}
\end{equation}
from this equations it follows the next expressions for time derivative of the scalar field $\dot{\phi}$ and the potential $V(\phi)$
\begin{equation}\label{eq17}
\dot{\phi}^2=\left(1+\omega_{\phi}\right)\rho_{\phi}   \ \ \,\ \ \  V(\phi)=\frac{1}{2}\left(1-\omega_{\phi}\right)\rho_{\phi}
\end{equation}
where the quintessence EoS parameter $\omega_{\phi}$  will be replaced by the holographic EoS  $\omega_{\Lambda}=\tilde{p_{\Lambda}}/\tilde{\rho_{\Lambda}}$ (with $\tilde{\rho_{\Lambda}}$ and $\tilde{p_{\Lambda}}$ given by Eqs. (\ref{eq7}),(\ref{eq8}) respectively). Therefore, as we expect from the holographic quintessence correspondence, the Eqs. (\ref{eq16}) will be written in terms of the holographic quantities. Taking into account that $\rho_{\Lambda}=3M_p^2H_0^2\tilde{\rho_{\Lambda}}$ one can write the first Eq. (\ref{eq17}) as
\begin{equation}\label{eq18}
\frac{d\phi}{dz}=\mp\sqrt{3}\frac{M_pH_0}{(1+z)H(z)}\left(\tilde{\rho_{\Lambda}}+\tilde{p_{\Lambda}}\right)^{1/2}
\end{equation}
where we have turned the time derivative to the redshift variable $z=1/a-1$. Substituting Eqs. (\ref{eq6} \ref{eq7},\ref{eq8}) into Eq. (\ref{eq18}) (with $H=H_0\tilde{H}$) gives
\begin{equation}\label{eq19}
\frac{d\phi}{dz}=\mp\frac{\sqrt{3}M_p}{1+z}\left[\frac{3\beta(2\alpha-3\beta)\Omega_{m0}+2(\alpha-1)(3\beta-2\alpha+2)C(1+z)^{2(\alpha-1)/\beta-3}}{6\beta\Omega_{m0}+3\beta(3\beta-2\alpha+2)C(1+z)^{2(\alpha-1)/\beta-3}}\right]^{1/2}
\end{equation}
where we used $\Omega_{r0}=0$, which is appropriate at low redshift. Despite the fact that this equation can be integrated exactly, the analytical expression is too large to be written here and instead we present the plot 
of $\phi$ as function of $z$ in Fig. 4.
\begin{center}
\includegraphics [scale=0.7]{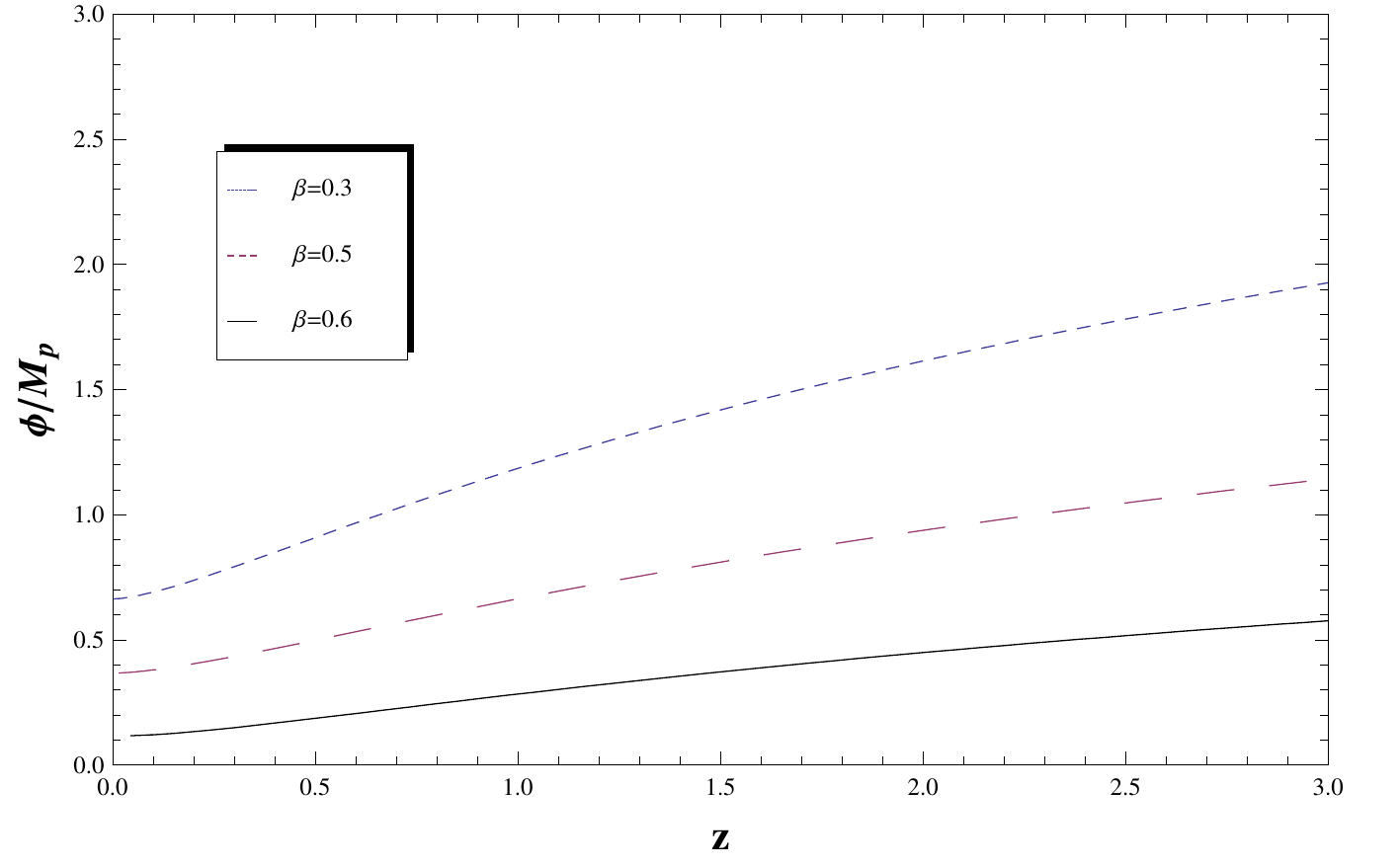}
\end{center}
\begin{center}
Figure 4: The holographic quintessence field $\phi$ versus redshift, according to table I.
\end{center}
where we have chosen the overall positive sign in Eq. (\ref{eq19}). Note that the field increases with $z$, but becomes finite at high redshift. This means that $\phi$ decreases as the universe expands. From the second equation in (\ref{eq16}) and proceeding as above, we obtain the expression for the potential in terms of the redshift $z$
\begin{equation}\label{eq20}
V(z)=\frac{3M_p^2H_0^2}{2}\left[\frac{3\beta-2\alpha}{2\alpha-3\beta-2}\Omega_{m0}(1+z)^3+\frac{2}{3}\frac{3\beta-\alpha+1}{\beta}C(1+z)^{2(\alpha-1)/\beta}\right]
\end{equation}
Fig. 5 shows the evolution of the quintessence potential with the redshift $Z$
\begin{center}
\includegraphics [scale=0.7]{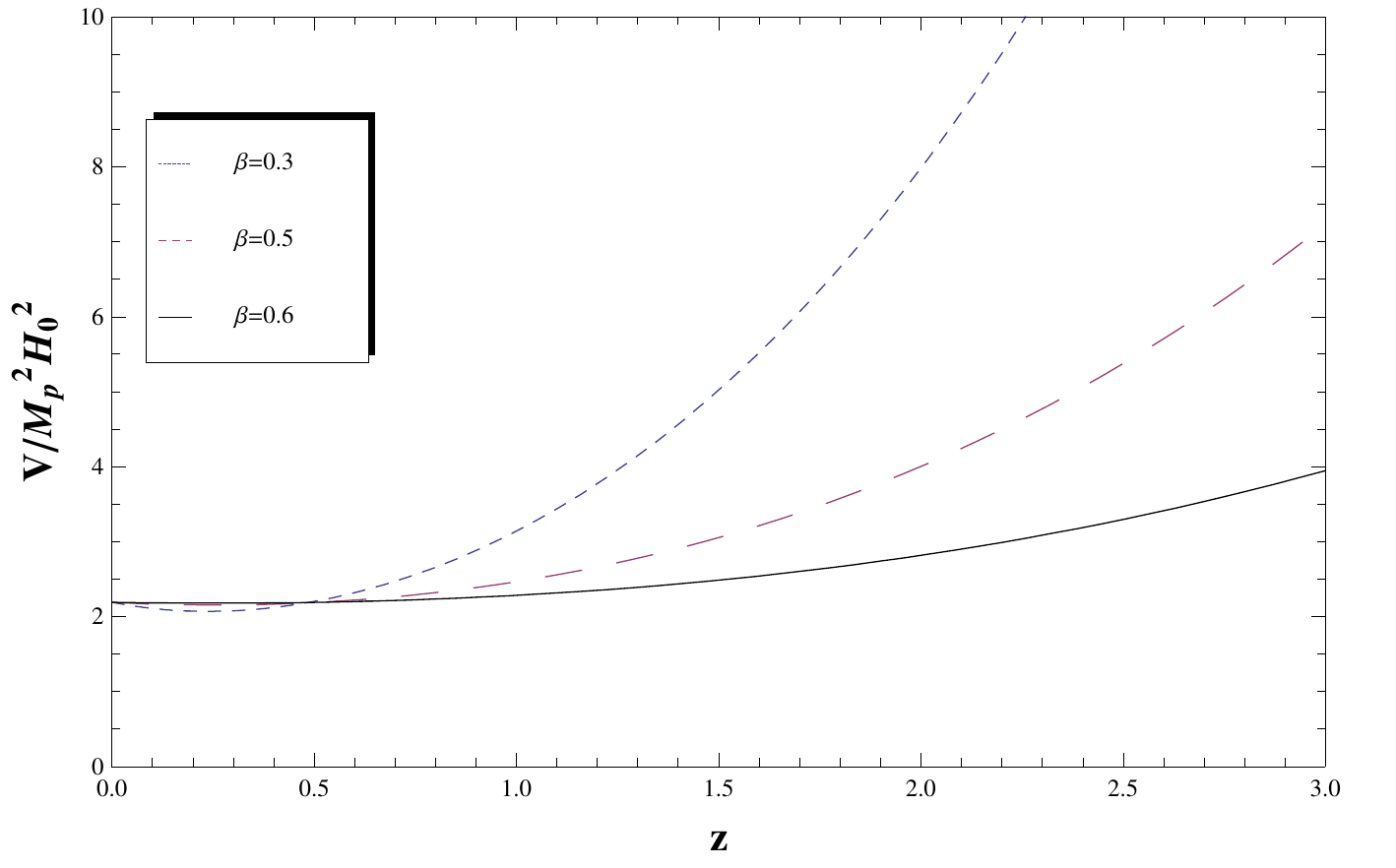}
\end{center}
\begin{center}
Figure 5: The evolution of the  quintessence potential versus redshift, according to table I.
\end{center}
Note that this potential decreases as the universe expands. Similar behavior has been obtained in \cite{jingfei} for an holographic tachyon model.
Numerically evaluating the above equations we can plot the potential dependence on the scalar field, as shown in Fig.6
\begin{center}
\includegraphics [scale=0.7]{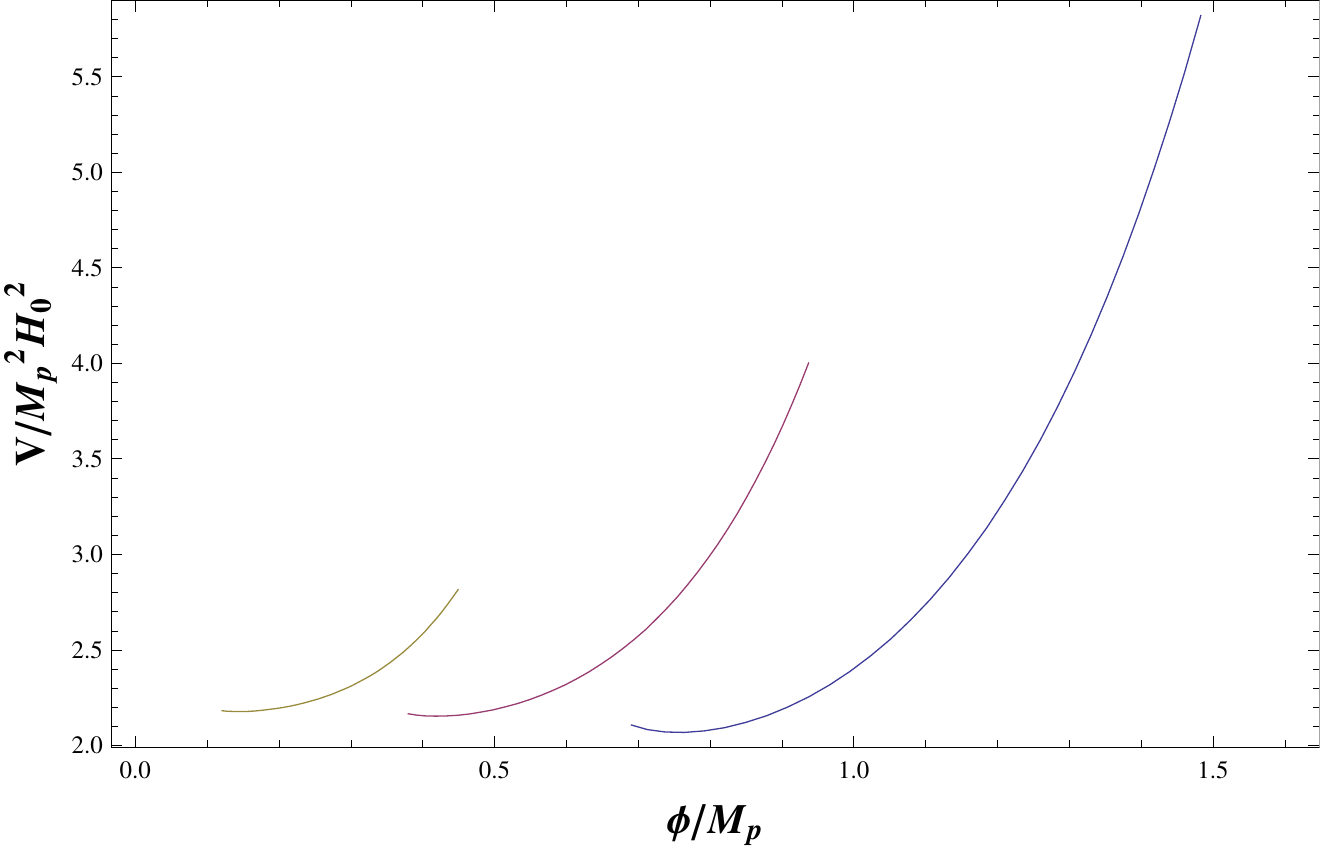}
\end{center}
\begin{center}
Figure 6: The holographic quintessence potential $V$ versus $\phi$, according to table I. From left to right $\beta=0.6,0.5,0.3$
\end{center}
Note that all the potentials are more steep in the early epoch, tending to be flat near today. Consequently, the quintessence field rolls down potential more slowly as the universe expands and the EoS parameter tends to negative values according to (\ref{eq16}), as $\dot{\phi}\rightarrow 0$.
We can also evaluate the potential taking the negative sign in Eq. (\ref{eq18}) and choosing the initial value for the quintessence field $\phi_0=M_p$ at the present. This results in the shift of the
value of the scalar field, but it does not change the shape of the potential (is shifted horizontally) and has no influence on the cosmological evolution. The behavior of the quintessence field in this case is seen from Fig. 4 by changing the sign of $\phi$. Therefore, the scalar field increases as the universe expands. Fig.7 shows the potential in this case
\begin{center}
\includegraphics [scale=0.7]{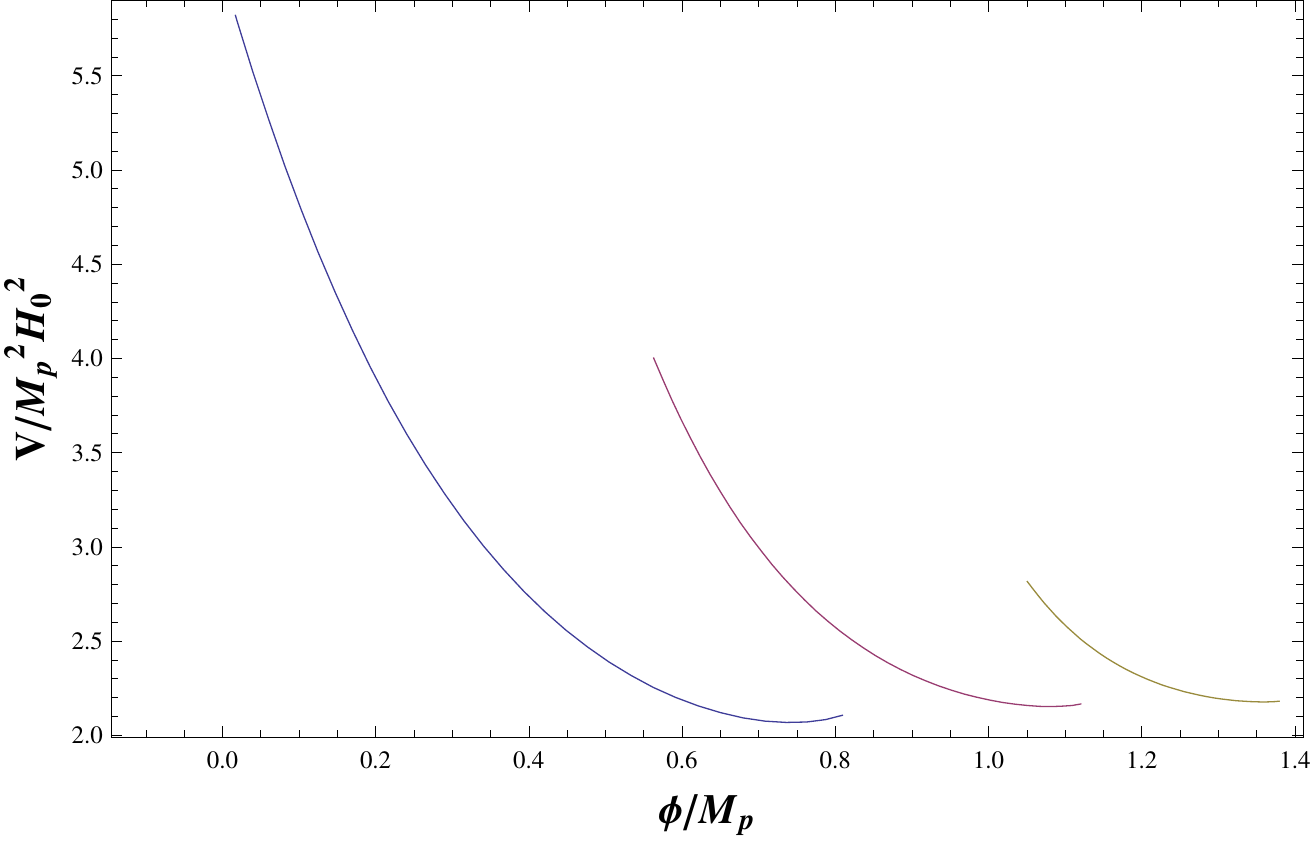}
\end{center}
\begin{center}
Figure 7: The holographic quintessence potential $V$ versus $\phi$, according to table I. In this case the negative option in Eq. (\ref{eq19}) has been taken. From left to right $\beta=0.3,0.5,0.6$
\end{center}
Note that the potentials are of a runaway type, and decrease as the universe expands. Similar behavior was obtained for quintessence potentials in (\cite{zong}).

\subsection*{The $\alpha>1$ case}
Following the same procedure as with $\alpha<1$, we will obtain the scalar field and  the shape of the potential that realizes the accelerated expansion for the case $\alpha>1$, taking the data from table II. Integrating the Eq. (\ref{eq19}), according to the values given in table II, Fig. 8 shows the quintessence field 
\begin{center}
\includegraphics [scale=0.7]{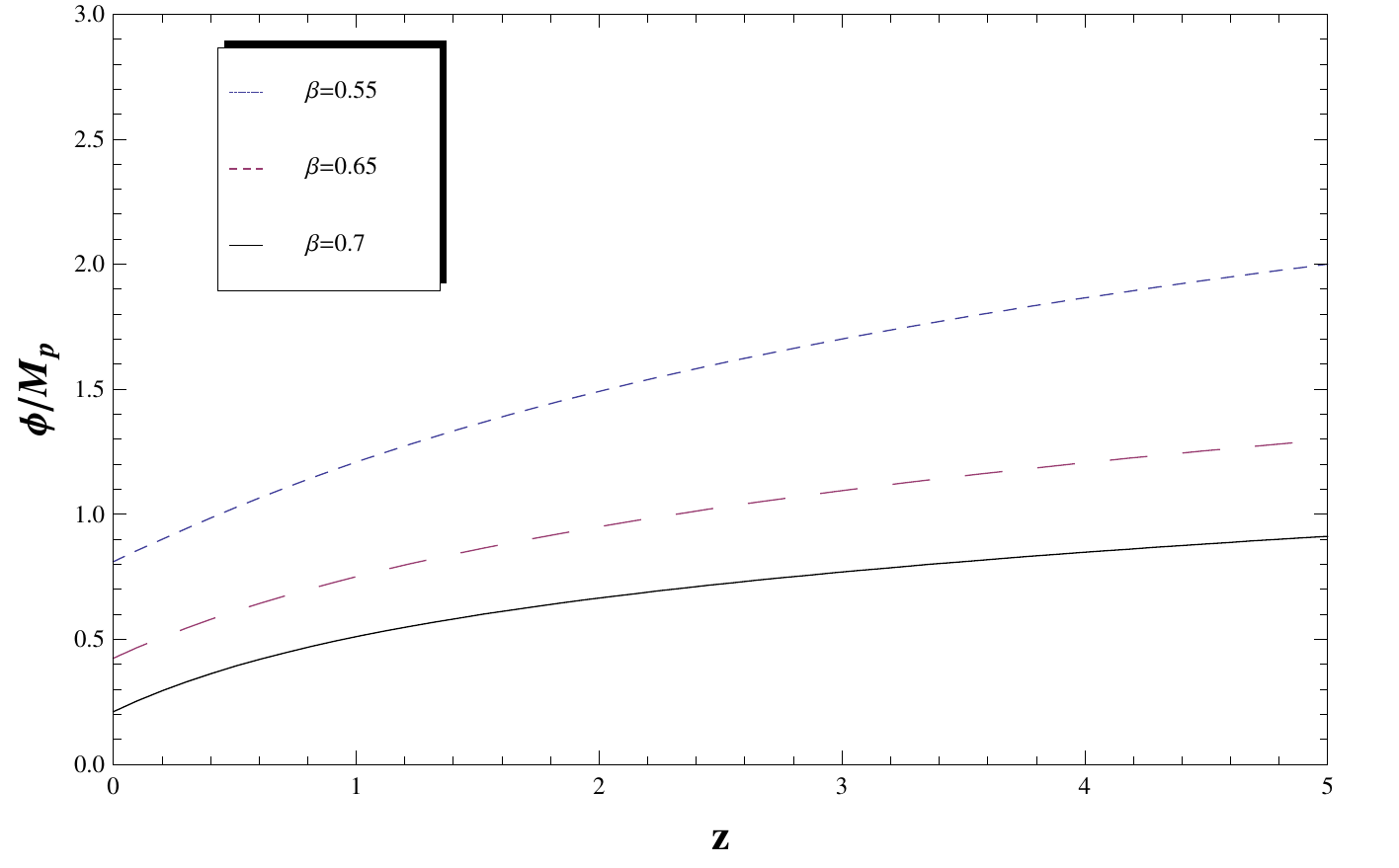}
\end{center}
\begin{center}
Figure 8: The evolution of the holographic quintessence field $\phi$, according to table II.
\end{center}
Fig. 9 shows the evolution of the quintessence potential with the redshift, according to the data given in table II.
\begin{center}
\includegraphics [scale=0.7]{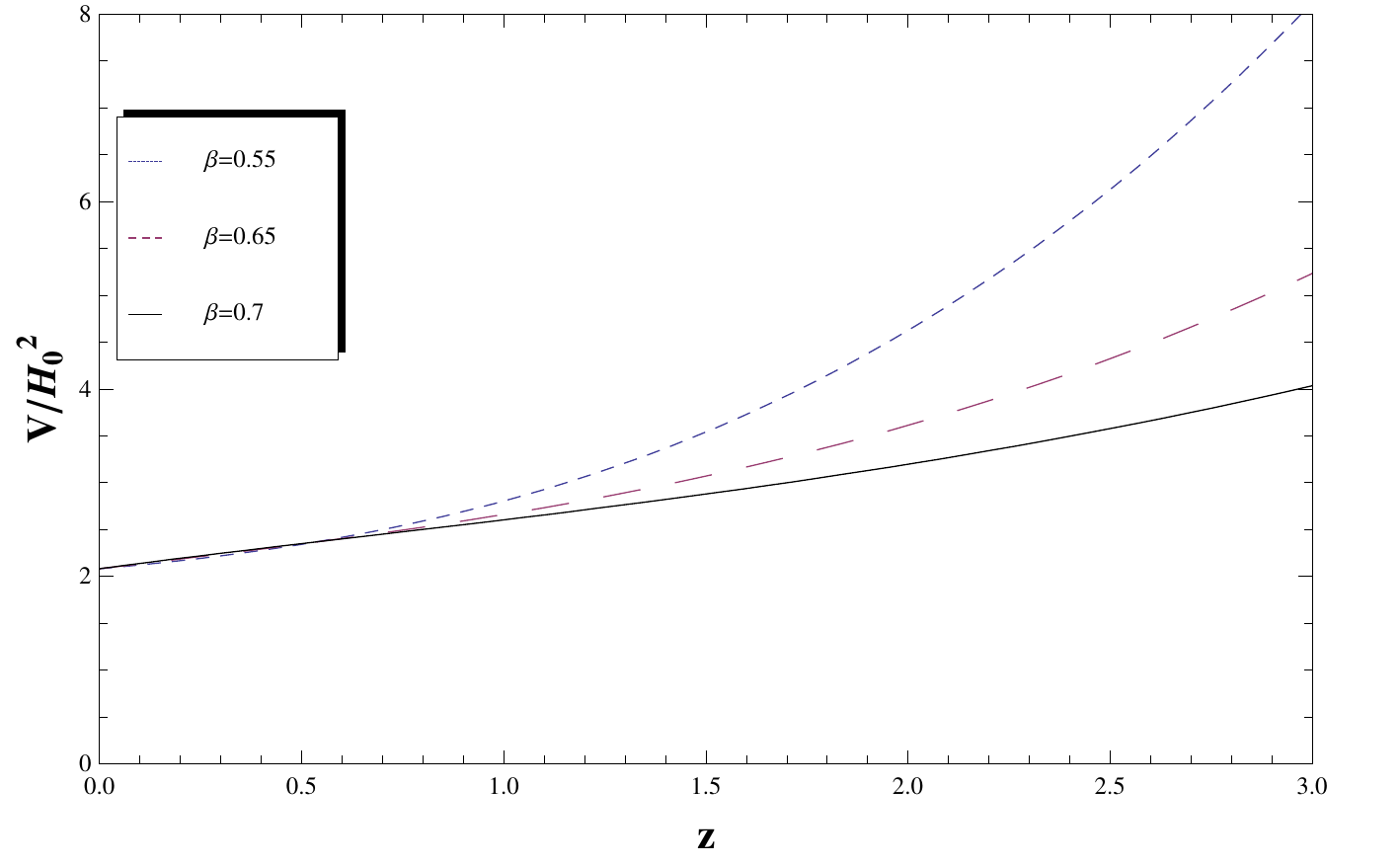}
\end{center}
\begin{center}
Figure 9: The evolution of the holographic quintessence potential $V$ in terms of the redshift, according to table II.
\end{center}
Fig. 10 shows the behavior of the quintessence potential with respect to the field, taking the data from table II.
\begin{figure}[hbtp]
\includegraphics [scale=0.7]{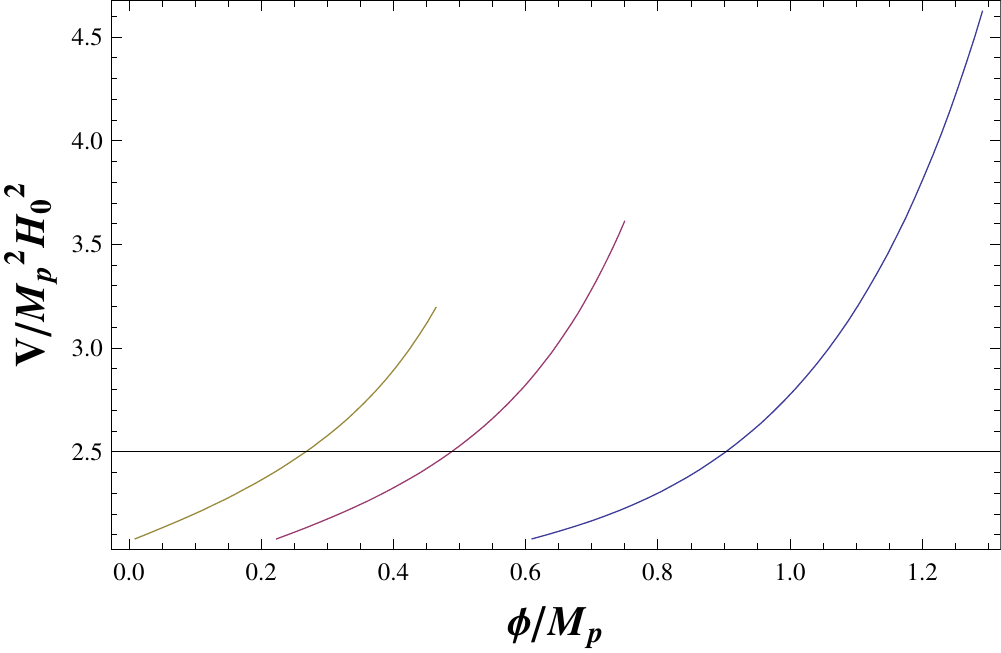}
\includegraphics [scale=0.7]{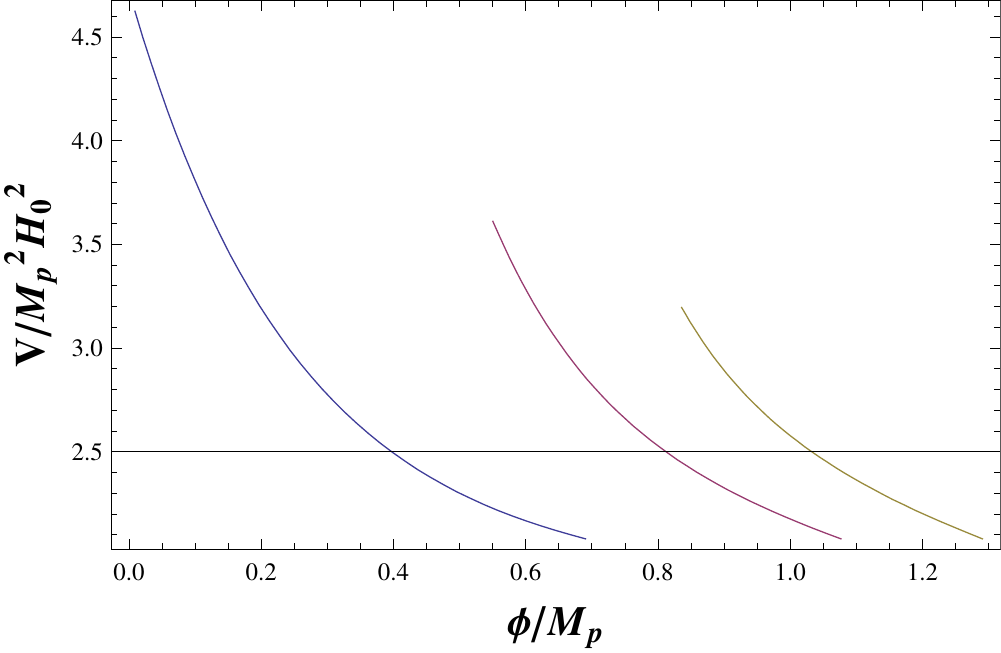}
\begin{center}
Figure 10: The behavior of the quintessence potential $V$ in terms of the scalar field, according to table II. The left plot corresponds to (+) in Eq. (\ref{eq19}), and from left to right $\beta=0.7,0.65,0.55$. The right plot corresponds to (-) sign in Eq. (\ref{eq19}) and from left to right $\beta=0.55,0.65,0.7$.
\end{center}
\end{figure}
Note all the quintessence dynamics behaves the same way as in the $\alpha<1$ case.
\section{Reconstruction of the tachyon model}
The tachyon models as a source of dark energy with different potential forms, have been discussed widely in the literature \cite{bagla,ying,liddle,copeland4}. 
The tachyon may be described by an effective field theory corresponding to some sort of tachyon condensate of string theory with an effective action in a gravitational background given by \cite{roo}, \cite{sen}
\begin{equation}\label{eq21}
S=\int d^4x\sqrt{-g}\left[\frac{R}{16\pi G}-V(\phi)\sqrt{1+g^{\mu\nu}\partial_{\mu}\phi\partial_{\nu}\phi}\right]
\end{equation}
where $V(\phi)$ is the tachyon potential and $R$ is the scalar curvature. The above action is considered to describe the
physics of tachyon condensation for all values of $\phi$ as long as string coupling and the second
derivative of $\phi$ are small.
Variating the second term in \ref{eq19} with respect to the metric we obtain the energy-momentum tensor of the tachyon field
\begin{equation}\label{eq22}
T_{\mu\nu}=-V(\phi)\sqrt{1+g^{\mu\nu}\partial_{\mu}\phi\partial_{\nu}\phi}g_{\mu\nu}+\frac{V(\phi)\partial_{\mu}\phi\partial_{\nu}\phi}{\sqrt{1+g^{\mu\nu}\partial_{\mu}\phi\partial_{\nu}\phi}}
\end{equation}
In the flat FRW background the energy density $\rho$ and the pressure density $p$ obtained from (\ref{eq20}) are given by
\begin{equation}\label{eq23}
\rho_T=-T_{0}^0=\frac{V(\phi)}{\sqrt{1-\dot{\phi}^2}}  \ \ \,\ \ \   p_T=T_i^i=-V(\phi)\sqrt{1-\dot{\phi}^2}
\end{equation}
From Eqs. (\ref{eq23}) follows the tachyon equation of state parameter 
\begin{equation}\label{eq24}
\omega_T=\dot{\phi}^2-1
\end{equation}
We see that no matter the potential, the tachyonic scalar field can not realize the equation of state crossing -1. Assuming the holographic nature to the tachyon, we should identify $\omega_T$ with $\omega_{\Lambda}$ and the tachyon energy density with that of the holographic model. Then $\dot{\phi}^2=1+\omega_{\Lambda}$, and turning to the redshift variable $z=1/a-1$, it follows 
\begin{equation}\label{eq25}
\frac{d\phi}{dz}=\mp\frac{1}{1+z}\left[\frac{1+\omega_{\Lambda}}{\tilde{H}^2}\right]^{1/2}\frac{1}{H_0}
\end{equation}
where $\omega_{\Lambda}$ (taking $\Omega_{r0}=0$ in Eqs. (\ref{eq7}) and (\ref{eq8}) is given by
\begin{equation}\label{eq26}
\omega_{\Lambda}=\frac{(2\alpha-3\beta-2)^2C(1+z)^{2(\alpha-1/\beta}}{3\beta\left[(3\beta-2\alpha)\Omega_{m0}(1+z)^3+(2\alpha-3\beta-2)C(1+z)^{2(\alpha-1)/\beta}\right]}
\end{equation}
and $\tilde{H}$ is given by Eq. (\ref{eq6}). From Eq. (\ref{eq26}) it follows that at high redshifts, the equation of state for the tachyon field approaches that of a pressure-less fluid and $\dot{\phi}^2\rightarrow 1$, avoiding the coincidence problem. For our model of holographic density, only
the holographic evolution for the cases $\alpha>1$ can be described by the tachyon. 
Fig. 9 shows the behavior of the holographic tachyon field. The integral can not be evaluated exactly, but we can  make a plot for a given interval
if we  integrate the Eq. (\ref{eq25}) from $z=0$ to a given value of $z$. This is equivalent to a displacement in $\phi$ by a constant value $\phi_0=\phi(z=0)$, which does not change the shape of the field itself. The evolution of the tachyon field for the data given in table II, is shown in Fig. 11. Note that the tachyon field decreases as the universe expands. Similar behavior has been obtained in \cite{jingfei}.
\begin{center}
\includegraphics [scale=0.7]{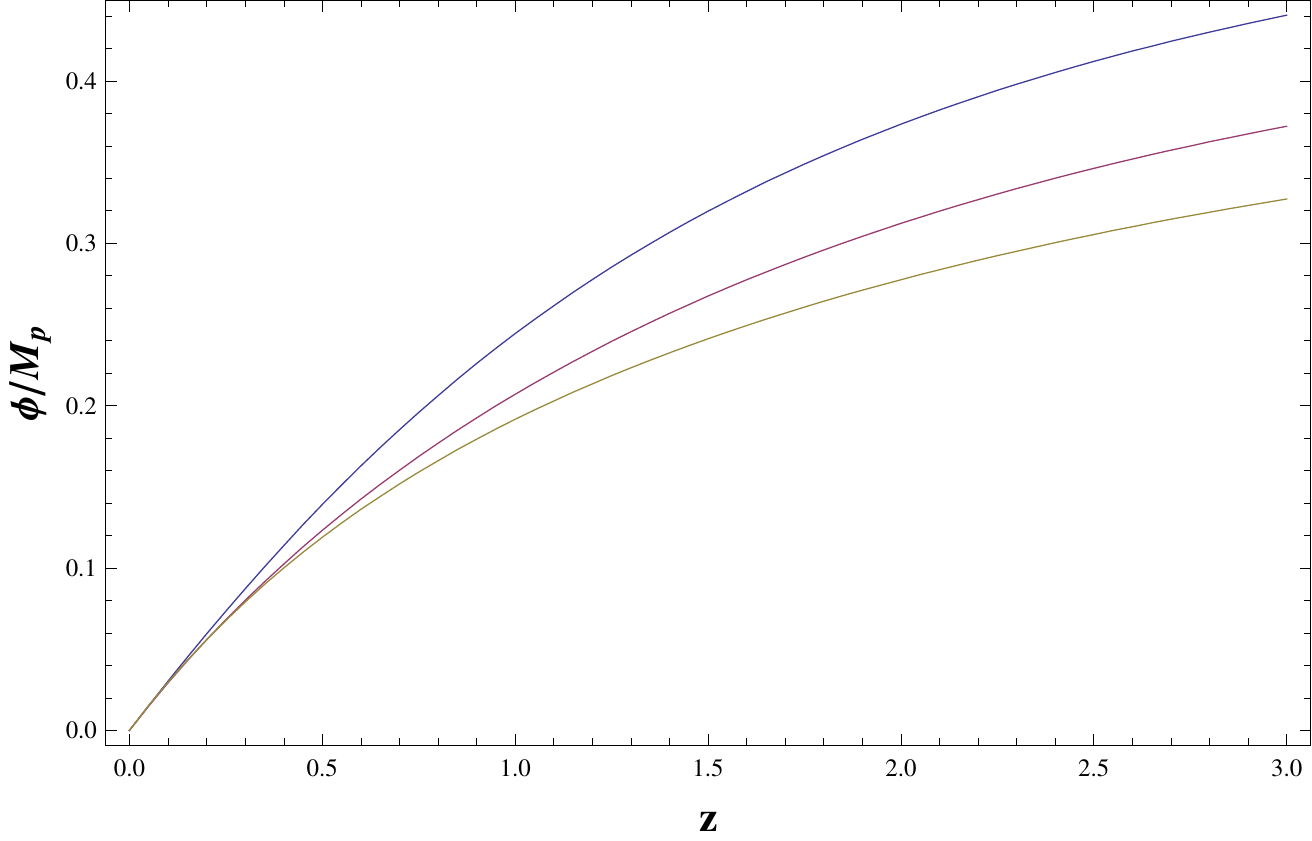}
\end{center}
\begin{center}
Figure 11: The behavior of tachyon field with the redshift, according to table II.
\end{center}
From Eqs. (\ref{eq23}) it is obtained $\rho_T p_T=-V^2(\phi)$, which reproduces the Chaplygin gas equation of state for constant $V(\phi)$. Assuming the holographic nature of the tachyon field, we replace $\rho_T$ and $p_T$ by the corresponding holographic quantities $\rho_{\Lambda}$ and $p_{\Lambda}$ from Eqs. (\ref{eq7}) and (\ref{eq8}), and obtain the following expression for the tachyon potential in terms of the redshift
\begin{equation}\label{eq27}
V(z)=3M_p^2H_0^2\left[\frac{2\alpha-3\beta}{3\beta}\Omega_{m0}C(1+z)^{(2\alpha+3\beta-2)/\beta}+\frac{3\beta-2\alpha+2}{3\beta}C^2(1+z)^{(\alpha-1)/\beta}\right]^{1/2}
\end{equation}
where we have set $\Omega_r=0$. Note the simplicity of the analytical form of the tachyon potential, which is displayed in figure 12.
\begin{center}
\includegraphics [scale=0.7]{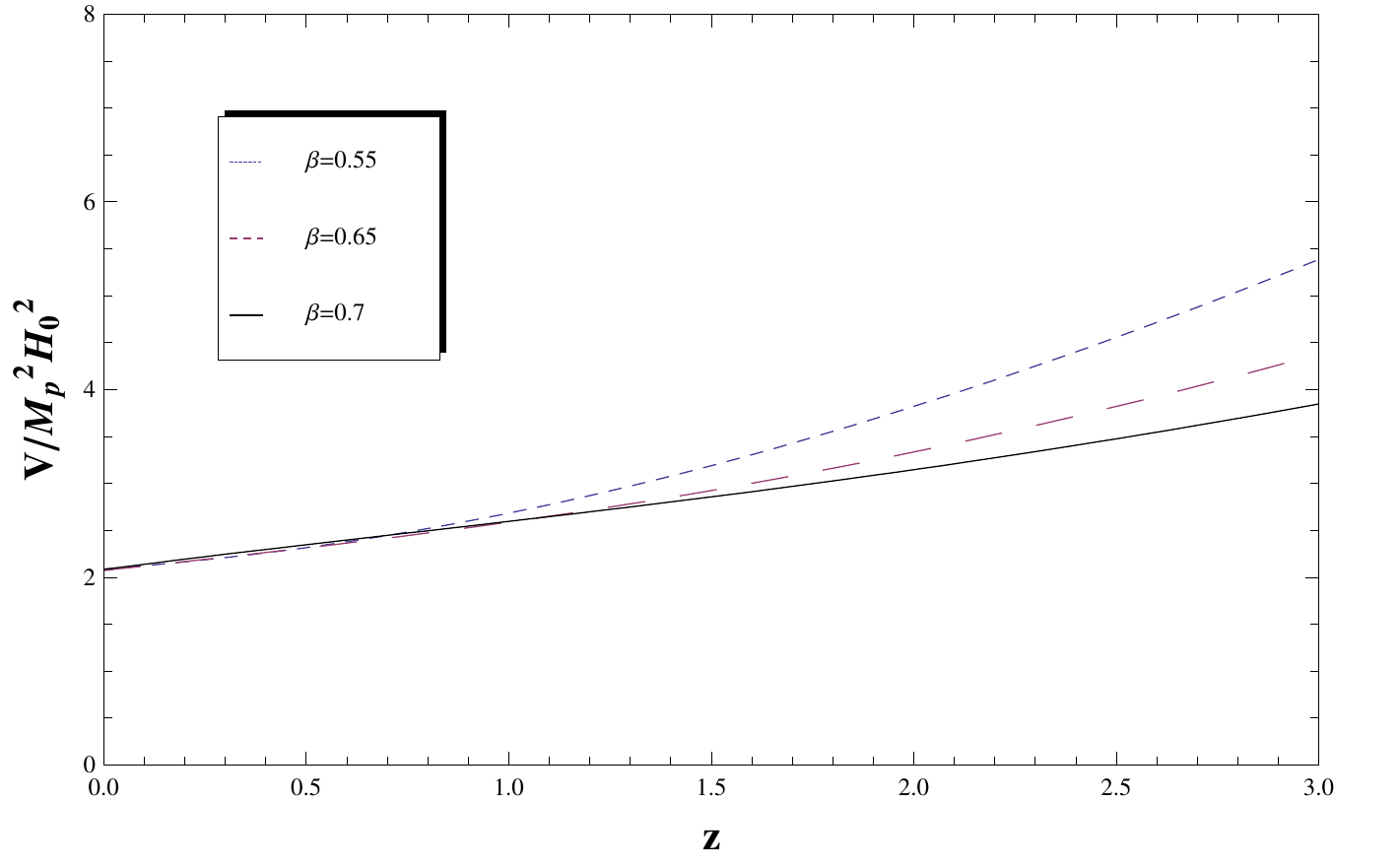}
\end{center}
\begin{center}
Figure 12: The behavior of tachyon potential with the redshift, according to table II.
\end{center}
The analytical form of the potential $V$ in terms of the tachyon field, cannot be obtained
due to the complexity of the tachyon field (\ref{eq25}), but we can obtain the holographic tachyon potential numerically, as shown in Fig. 13 for both signs in Eq. (\ref{eq25}). Note that in both cases the potential decreases as the universe expands. 
\begin{figure}[h]
\begin{center}
\includegraphics [scale=0.7]{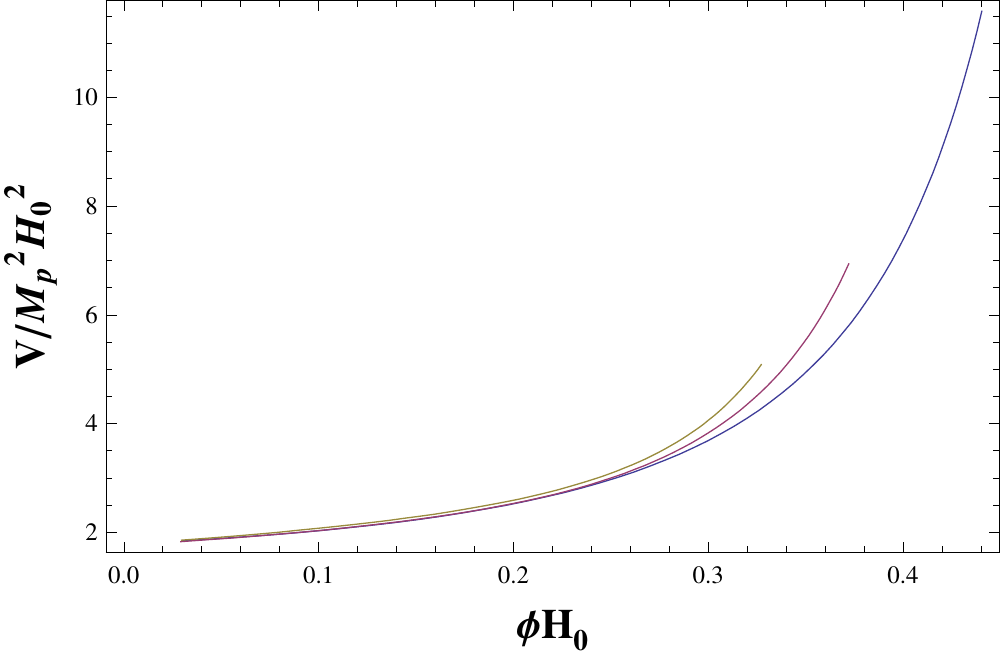}
\includegraphics [scale=0.7]{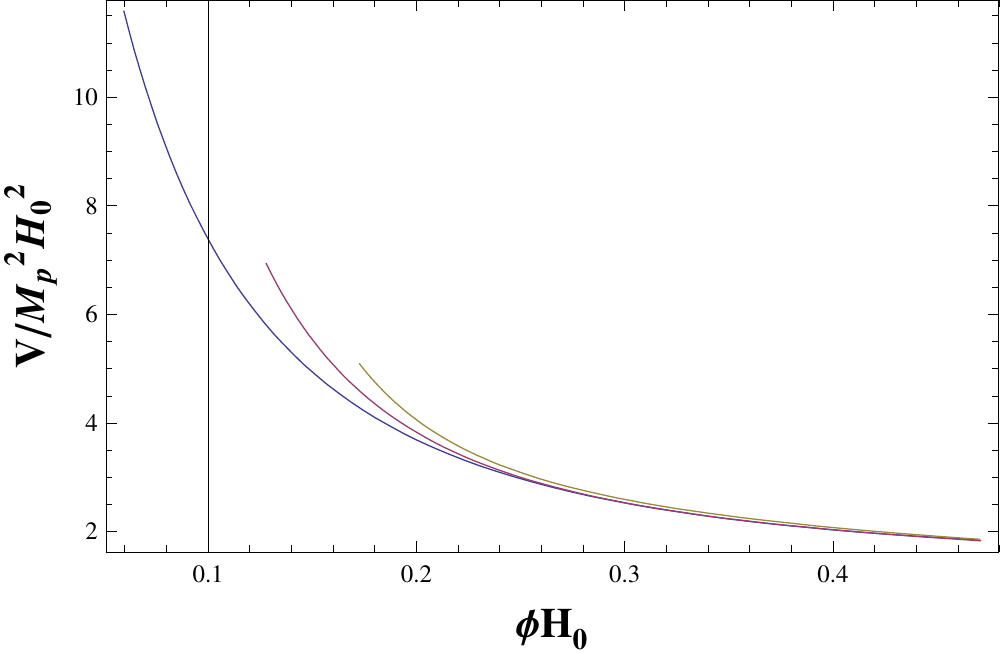}
\end{center}
\begin{center}
Figure 13: The behavior of the tachyon potential $V$ in terms of the tachyon field, according to table II. The left plot corresponds to (+) in Eq. (\ref{eq25}), and from left to right $\beta=0.7,0.65,0.55$. The right plot corresponds to (-) sign in Eq. (\ref{eq25}) and from left to right $\beta=0.55,0.65,0.7$.
\end{center}
\end{figure}

\section{Discussion}
We have carried out a detailed analysis about the cosmological evolution of the holographic quintessence and tachyon models of dark energy, in the frame of the infrared cut-off proposal for the holographic density  given in (\ref{eq2}). The only parameter in this model of holographic dark energy which needs to be fitted by the observational data is $\beta$ and is positive in all cases. Thus the parameter $\alpha$ (with $\beta$ being positive) plays a significant
role in the holographic evolution of the universe. When $\alpha<1$, the holographic evolution will make the equation of state cross $\omega = -1$ (from $\omega > -1$ evolves to $\omega < -1$). If $\alpha>1$, the the EoS parameter will stay in the region $-1<\omega<0$.
In the case of quintessence field, we have carried out the reconstruction of the field and the potential using the correspondence with the holographic principle, for both cases of $\alpha$: for $\alpha<1$ we reconstructed the holographic  quintessence model in the region before the crossing, i.e. for $\omega>-1$ which is the allowed region for the quintessence field. For $\alpha>1$ the reconstruction was successful in all the region, since in this case $\omega_{\Lambda}>-1$ as seen in Fig. 3 for the table II. From figures
4-7, we can see the dynamics of the quintessence field explicitly for the case $\alpha<1$. In Fig. 6 the potential is more steep in the early times, becoming flat at the present epoch and hence the kinetic term is gradually decreasing as obtained in \cite{xin} for the holographic quintessence based on the future event horizon cut-off. The behavior of the potential shown in Fig. 7 for the minus sign in the solution of Eq. (\ref{eq19}), resembles that presented in work \cite{zong}. See also \cite{dragan} and \cite{martins} for the reconstruction of the quintessence potential with similar results. In both cases of Figs. 6 and 7 the potential decreases as the universe expands as is shown in Fig. 5. The reconstruction for the case $\alpha>1$ gives completely similar results as can be judged by the Figs. 8-10.\\
The connection between the tachyon and the holographic dark energy has been established for the second set of parameters given in table II, for $\alpha>1$ (the values of table I lead to imaginary $\phi$) . The holographic tachyon model has been constructed according to equations (\ref{eq25})-(\ref{eq27}). The dynamics of the holographic tachyon is shown in Figs. 11-13. Comparing with the case of quintessence, the tachyon potential shows very similar behavior, as seen from figs. 9 and 12 , decreasing with time and flattening in the near epoch, which lead to an EoS parameter close to $-1$ in the future. The tachyon fields (with (+) sign in Eq. (\ref{eq25}) and potentials show the same behavior as the ones presented in \cite{jingfei}. The potential for the field of the opposite sign in Eq. (\ref{eq25}), as shown in Fig. 13, has been studied in \cite{ying} using different parametrizations of the equation of state parameter for the tachyon field.
In summary, it can be seen that the reconstruction has been successful in reproducing the main features of the potentials for quintessence and tachyon models according to the results reported in the literature.
One should realize, nevertheless, that despite the fact that this holographic model which depends on local quantities, gives rise to the reconstruction in a direct and unambiguous way, this reconstruction
is settled at the phenomenological level, as the theoretical root of the holographic density still to be investigated.
However, the obtained results may suggest that we are in the appropriate way to understand one of the intriguing problems of the modern science.

\section*{Acknowledgments}
This work was supported by the Universidad del Valle, under project CI-7713.

\end{document}